# Lack of dispersion cancellation with classical phase-sensitive light

J.D. Franson

*Physics Department, University of Maryland, Baltimore County, Baltimore, MD 21250*
(Dated: September 11, 2009)

J.H. Shapiro recently argued that nonlocal dispersion cancellation using entangled pairs of photons is essentially classical in nature, based on a comparison with a classical model in which two stationary, chaotic beams of light have phases and frequencies that are anti-correlated, which he refers to as "phase-sensitive" light (arXiv:0909.2514). It is shown here that there is no physical cancellation of dispersion for classical light of that kind, and Shapiro's results merely reflect the fact that identical dispersion occurs in both beams. The origin of the cross-correlations between the intensities of the two beams is shown to be completely different in the classical and quantum-mechanical cases.

## I. INTRODUCTION

In nonlocal dispersion cancellation [1-5], a pair of energy-time entangled photons remain coincident even after they have passed through two distant dispersive media, as illustrated in Fig. 1. J.H. Shapiro [6] has recently considered a classical model in which the cross-correlation between the intensities of two beams of classical light (the usual peak in the Hanbury-Brown and Twiss effect) is also unaffected by two distant dispersive media. Although Shapiro accepts the fact that there is no classical theory that can completely agree with nonlocal dispersion cancellation for entangled photons, he nevertheless argues that the origin of nonlocal dispersion cancellation is essentially classical in nature. It will be shown here that there is no physical cancellation of dispersion in the classical example considered by Shapiro, and that the origin of the classical effects is completely different from the quantum-mechanical case.

The essential feature of nonlocal dispersion cancellation is that entangled pairs of photons will be detected at exactly the same time (in the limit of large bandwidths and low intensities), despite the presence of the dispersive media [1]. In classical electromagnetism, the only way to produce detector outputs that are always coincident would be to emit very short pulses at correlated times. But the width of the classical pulses would be greatly increased due to dispersion in the two media, regardless of the sign of the dispersion coefficients, and it seems apparent that there can be no classical model that reproduces or explains the results of nonlocal dispersion cancellation [3].

Shapiro [6] considered two stationary, chaotic Gaussian beams of light whose phases and frequencies are anti-correlated, which he refers to as "phase-sensitive" beams. He showed that the cross-correlation $\langle I_1(t)I_2(t')\rangle$ between the intensities $I_1(t)$ and $I_2(t)$ of the two beams is unaffected by the presence of the two dispersive media, as illustrated in Fig. 2. The fact that the width of the peak in the cross-correlation function is unaffected by the presence of the dispersive media may appear to be analogous to nonlocal dispersion cancellation for entangled photons, and Shapiro concluded that "their physical origins are *identical* and essentially *classical*".

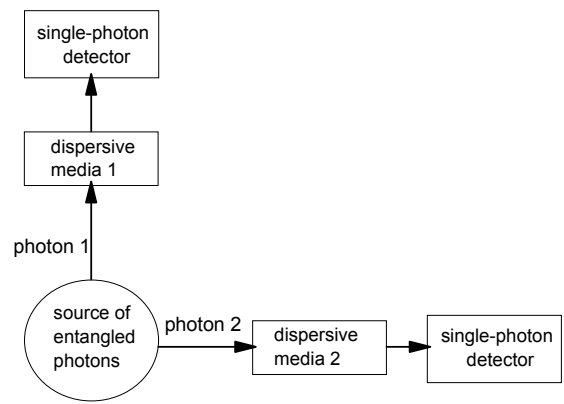

Fig. 1. Nonlocal dispersion cancellation [1-5], in which a pair of energy-time entangled photons propagate through two distant dispersive media. The dispersion in one medium can be cancelled out nonlocally by the dispersion in the other medium if the dispersion coefficients of the two media are equal and opposite. (From Ref. [3].)

Although the width of the peak in Fig. 2 is unaffected by dispersion, the origin of this classical effect is totally different from the quantum-mechanical case. As will be described in more detail in Section II, the electric field in one of the classical beams is equal to the complex conjugate of that in the other beam. As a result, the two beams only differ in their relative phase and they have identical intensities as illustrated in the upper half of Fig. 3. Because the two beams have identical intensities, their cross-correlation function corresponds to the usual Hanbury-Brown and Twiss effect for chaotic light.

It will be shown here that there is no physical cancellation of dispersion in Shapiro's example and that the intensities of both beams are modified as a result of dispersion, as illustrated in the lower half of Fig 3. The effects of dispersion are the same in the two beams, so that the intensities of the two beams remain identical and correlations such as $\langle I_1(t)I_2(t')\rangle$ are unchanged. These classical results are due to the local properties of the dispersive media and the fields within them, unlike the situation for nonlocal dispersion cancellation [1-5]. There is no physical cancellation of dispersion in this example, and Shapiro's results merely reflect the fact that the dispersion is the same in the two beams and will not affect statistical properties such as the difference of the two intensities, for example.

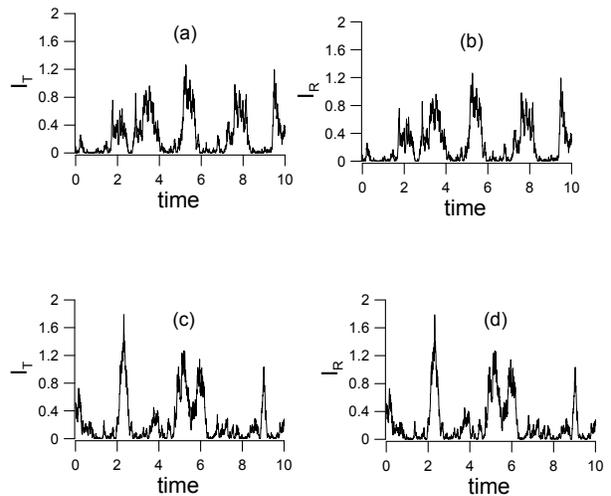

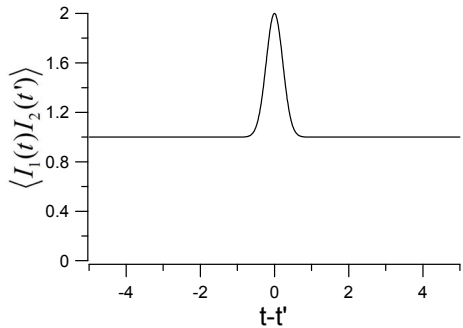

Fig. 2. Cross-correlation $\langle I_1(t)I_2(t')\rangle$ of the intensities of two chaotic classical beams of light with identical intensities as a function of the time delay $t-t'$ (arbitrary units). Shapiro [6] showed that the width of the peak near zero time delay (the usual Hanbury-Brown and Twiss effect) can be unaffected by the dispersion in two distant media, which he considered to be analogous to nonlocal dispersion cancellation. The constant background for large time delays produces a large number of detection events that are not coincident, and this model does reproduce the essential feature of nonlocal dispersion cancellation that makes it nonclassical.

As can be seen in Fig. 2, there is a broad background in the intensity correlation function in addition to the sharp peak near zero time delay. This allows two detectors to register counts at widely different times, which is totally different from what happens in nonlocal dispersion cancellation using entangled photons, where the detectors only register counts at the same time. Shapiro mentions this difference between the classical and quantum results but considers it to be insignificant when interpreting his results. As mentioned above, it is the fact that all of the photon pairs are coincident that makes it impossible to describe the quantum results by a classical model. Shapiro's model does not reproduce this essential feature of the quantum-mechanical results and it is not analogous to nonlocal dispersion cancellation.

The entangled nature of the photon pairs allows the dispersion coefficients of the two media to coherently cancel each other, thus physically eliminating the effects of dispersion altogether. This is not possible classically, and it will be seen that the origins of these effects are totally different, both mathematically and physically.

Fig. 3. Typical intensities of two classical beams of chaotic light with anti-correlated phases (arbitrary units). (a) Intensity of beam 1 with no dispersive medium. (b) Intensity of beam 2 with no dispersive medium. (c) Intensity of beam 1 after passing through a dispersive medium. (d) Intensity of beam 2 after passing through a dispersive medium with an equal but opposite dispersive coefficient. There is no cancellation of dispersion in this example; instead, both beams are dispersed in the same way, which maintains the statistical correlations between the intensities of the two beams. These plots were simulated in Mathematica using a Markov process. (Taken from Ref. [3].)

The "phase-sensitive" beams of light introduced by Shapiro [6] will be described in more detail in Section II. The cross-correlations and other statistical properties of these beams after they pass through two dispersive media will be derived in a straightforward way in Section III. The quantum-mechanical and classical results are compared in Section IV, where it is apparent that the origin of the insensitivity of the cross-correlations to dispersion is fundamentally different in the two cases. A summary and conclusions are provided in Section V.

## II. "PHASE-SENSITIVE" LIGHT

In his model for nonlocal dispersion cancellation, Shapiro considered two stationary, chaotic Gaussian beams of light whose phases and frequencies are anti-correlated, which he referred to as "phase-sensitive" beams [6]. His description of these fields was somewhat abstract and given in terms of their assumed statistical properties. Here we will consider a specific example of a classical field that has all of the properties assumed by Shapiro. This has the advantage of allowing a straightforward calculation of $\langle I_1(t)I_2(t')\rangle$ in the next section, as well as providing sufficient insight to determine whether or not nonlocal dispersion cancellation is essentially classical or not.

As noted by Shapiro [6], a chaotic field is equivalent to an incoherent mixture of coherent states. We can therefore write the complex classical electric field (analytic signal [7]) $E_1(t)$ emitted by source 1 in the form

$$E_1(t) = \sum_n c_n e^{i\phi_n} e^{-i\omega_0 t} e^{-i\Delta\omega_n t}. \quad (1)$$

Here the coefficients $c_n$ are real and chosen at random from a suitable probability distribution, and the phase shifts $\phi_n$ are randomly distributed. For simplicity, we consider a sum over a large number of discrete frequencies $\omega_n = \omega_0 + \Delta\omega_n$, where the frequency shifts $\Delta\omega_n$ are a function of the mode $n$ and $\omega_0$ is a reference frequency; an integral over a continuum of frequencies would give similar results.

The essential idea of Shapiro's classical model [6] is that the field emitted by source 2 has phases and frequencies that are anti-correlated with those of source 1, as is the case in the entangled state that gives rise to nonlocal dispersion cancellation. Thus the complex classical electric field $E_2(t)$ emitted by source 2 is assumed to have the form

$$E_2(t) = \sum_n c_n e^{-i\phi_n} e^{-i\omega_0 t} e^{i\Delta\omega_n t}. \quad (2)$$

Aside from an irrelevant phase factor of $\exp[2i\omega_0 t]$, it can be seen that

$$E_2(t) = E_1^*(t). \quad (3)$$

As a result, $E_1(t)E_2(t)$ is equal to the intensity of beam 1:

$$E_1(t)E_2(t) = E_1^*(t)E_1(t) = I_1(t) \quad (4)$$

and Eq. (39) of Shapiro's paper [6] is satisfied. This condition is the essential feature of his model, which he referred to as "phase-sensitive" states in the title of his paper; it is satisfied here as well. It also follows from Eq. (3) that the intensities of the two beams are identical (they only differ by a phase factor):

$$I_1(t) = I_2(t). \quad (5)$$

We will be interested in the product of the intensities $I_1(t)$ and $I_2(t')$ of the two classical beams, which can be written as

$$I_1(t)I_2(t') = \left(E_1^*(t)E_1(t)\right)\left(E_2^*(t')E_2(t')\right) = |E_1(t)E_2(t')|^2. \quad (6)$$

We will take an ensemble average later in the calculation after we have first considered the possible role of dispersion cancellation in the product of the two fields. The fact that the cross-correlation of the intensities can be related to the cross-correlation of the electric fields is what motivates Shapiro's choice of the "phase-sensitive" beams.

Combining Eqs. (1) through (6) gives

$$I_1(t)I_2(t') = \left|\left(\sum_n c_n e^{i\phi_n} e^{-i\omega_0 t} e^{-i\Delta\omega_n t}\right)\left(\sum_{n'} c_{n'} e^{-i\phi_{n'}} e^{-i\omega_0 t'} e^{i\Delta\omega_{n'} t'}\right)\right|^2. \quad (7)$$

It is important to note that Eq. (7) corresponds to a double sum over $n$ and $n'$, which will be seen to be the main difference between the classical calculation and the corresponding quantum calculation, which has only a single sum over one index $n$.

From Eq. (5), the intensities of the two beams are equal. As a result, the cross-correlation of their intensities is equal to the correlation of the intensity of either beam with itself:

$$\langle I_1(t)I_2(t')\rangle = \langle I_1(t)I_1(t')\rangle = \langle I_2(t)I_2(t')\rangle \quad (8)$$

This is also the case for the more general model described in Ref. [6]. Since the beams are assumed to be stationary and chaotic, all of the intensity correlations in Eq. (8) exhibit the well-known Hanbury-Brown and Twiss peak, which is a factor of two larger for zero time delay than it is for large time delays as illustrated in Fig. 2. While there is a narrow peak corresponding to nearly coincident detection events, there is also a large background corresponding to detection events that occur at uncorrelated times. It is important to note that this situation is very different from the quantum-mechanical results using entangled photons, as will be discussed in Section IV.

### III. EFFECTS OF DISPERSION

Eqs. (1) through (7) give the quantities of interest at the time the fields are emitted by the sources. The corresponding quantities after the fields have passed through the two dispersive media will be denoted by primes. For example, the field emitted by source 1 will be denoted $E_1'(t)$. The form of these fields can be obtained by expanding the wave vector $k(\omega)$ in each medium in a Taylor series expansion [1]:

$$\begin{aligned} k_1 &= k_0 + \alpha_1 \Delta\omega_1 + \beta_1 \Delta\omega_1^2 \\ k_2 &= k_0 - \alpha_2 \Delta\omega_1 + \beta_2 \Delta\omega_1^2. \end{aligned} \quad (9)$$

Here the coefficients $\alpha_i$ and $\beta_i$ are related to the group velocity and dispersion, respectively. Ignoring the group velocities, which only give a constant offset of the two beams, the product of the intensities after the dispersive media becomes

$$I_1'(t)I_2'(t') = \left| \left( \sum_n c_n e^{i\phi_n} e^{-i\omega_0 t} e^{-i\Delta\omega_n t} e^{i\beta_1 \Delta\omega_n^2 L} \right) \right.$$
$$\left. \times \left( \sum_{n'} c_{n'} e^{-i\phi_{n'}} e^{i\omega_0 t'} e^{i\Delta\omega_{n'} t'} e^{i\beta_2 \Delta\omega_{n'}^2 L} \right) \right|^2 \quad (10)$$

where $L$ is the length of the media.

It is instructive to consider a typical cross-term $T_{nn'}$ in the product of Eq. (10) corresponding to arbitrary indices $n$ and $n'$:

$$T_{nn'} = e^{-i\omega_0(t+t')} c_n c_{n'} e^{i(\phi_n - \phi_{n'})} e^{-i(\Delta\omega_n t - \Delta\omega_{n'} t')} e^{i(\beta_1 \Delta\omega_n^2 + \beta_2 \Delta\omega_{n'}^2)L} \quad (11)$$

where $n \neq n'$ in general. It is important to note that $\Delta\omega_n \neq \Delta\omega_{n'}$ in general, and as a result the effects of the dispersive media do not cancel out even if $\beta_1 = -\beta_2$, since

$$(\beta_1 \Delta\omega_n^2 + \beta_2 \Delta\omega_{n'}^2) \neq 0. \quad (12)$$

In fact, it can be seen that both fields undergo dispersion:

$$E_1'(t) = \sum_n c_n e^{i\phi_n} e^{-i\omega_0 t} e^{-i\Delta\omega_n t} e^{i\beta_1 \Delta\omega_n^2 L} \neq E_1(t)$$
$$E_2'(t) = \sum_{n'} c_{n'} e^{-i\phi_{n'}} e^{-i\omega_0 t} e^{i\Delta\omega_{n'} t} e^{i\beta_2 \Delta\omega_{n'}^2 L} \neq E_2(t). \quad (13)$$

Independent dispersive effects occur locally in both beams, and there is obviously no interaction between the two fields or any physical cancellation of dispersion.

Nevertheless, it can be seen that

$$E_2'(t) = E_1'^*(t) \quad (14)$$

if $\beta_1 = -\beta_2$, just as in Eq. (3). This can be used to write the product of the intensities as

$$I_1'(t)I_2'(t') = |E_1'(t)E_2'(t')|^2 = |E_1'(t)E_1'^*(t')|^2 = I_1'(t')I_1'(t). \quad (15)$$

Physically, this means that the intensities of both beams were identical initially from Eq. (3) and from Eq. (14) they remain identical after passing through the dispersive media. Thus the intensities of both beams undergo identical dispersion, as illustrated schematically in Fig. 3 [3].

The classical electric fields $E_1(t)$ and $E_1'(t)$ differ only in that the random phase $\phi_n$ is replaced by $\phi_n + \beta_1 \Delta\omega_n^2 L$, which is also a random phase. Thus all of the statistical properties of the fields are unchanged by the dispersion, so that

$$\langle I_1(t) \rangle = \langle I_1'(t) \rangle$$
$$\langle I_1(t)I_1(t') \rangle = \langle I_1'(t)I_1'(t') \rangle \quad (16)$$

A more formal proof of this has been given by Wang et al. [8]. Combining Eqs (15) and (16) shows that

$$\langle I_1(t)I_2(t') \rangle = \langle I_1'(t)I_2'(t') \rangle. \quad (17)$$

Thus the fourth-order correlations between the two fields are unaffected by the dispersion, as was shown by Shapiro using a different method [6].

The dependence of the intensity correlation on the dispersive coefficients $\beta_1$ and $\beta_2$ can be obtained by once again expanding in a Taylor series:

$$\langle I_1(t)I_2(t) \rangle - \langle I_1'(t)I_2'(t) \rangle = a + b_1\beta_1 + b_2\beta_2 + c_1\beta_1^2 + c_2\beta_2^2 + d\beta_1\beta_2. \quad (18)$$

Here the group velocity terms can be ignored as before and Eq. (9) already assumes that the dispersion is small, so that any higher-order terms can be neglected as well. The fact that the left-hand side of Eq. (18) goes to zero for $\beta_1 = -\beta_2$ and is positive otherwise restricts the coefficients in the expansion in such a way that

$$\langle I_1(t)I_2(t') \rangle - \langle I_1'(t)I_2'(t') \rangle = c_1(\beta_1 + \beta_2)^2. \quad (19)$$

Eq. (19) is similar in appearance to the quantum-mechanical result [1], but it is apparent from Eq. (11) that this is not due to any physical cancellation of the dispersion in the propagation of the fields, unlike the quantum-mechanical case. The form of Eq. (19) merely reflects the fact that the intensities of the two beams remain identical if $\beta_1 = -\beta_2$.

The lack of physical dispersion cancellation in the classical model can be further illustrated by considering the case in which the coefficients $c_n$ in Eq. (1) are chosen in such a way that the classical field of beam 1 corresponds to a short pulse of chaotic light. Using the same coefficients in Eq. (2) will maintain the condition that $E_2(t) = E_1^*(t)$ and $E_2'(t) = E_1'^*(t)$, which is the essential feature of the "phase-sensitive" fields referred to in the title of Shapiro's paper. But in this case, both pulses will undergo a large amount of dispersion and will be greatly broadened [3]. This shows that the "phase-sensitive" nature of the fields does not eliminate the effects of dispersion. In the case of stationary fields as opposed to short pulses, confining our attention to statistical properties such as the difference in the intensities of the two beams provides an illusion that the dispersion is being eliminated.

In an earlier paper, Torres-Company et al. [9] considered a similar classical model in which two stationary chaotic beams were in identical states, such as that of Eq. (1). They showed that the fourth-order coherence of the two beams would be maintained if $\beta_1 = \beta_2$ in that case. Once again, both beams undergo identical dispersion and there is no cancellation of dispersion. The model considered here is similar to that of Ref. [9] in that the two beams have identical intensities initially and undergo identical dispersion processes in both cases. It would be more correct to refer to phenomena of this kind as "identical dispersion" rather than dispersion cancellation [3].

## IV. COMPARISON OF THE CLASSICAL AND QUANTUM RESULTS

In the original quantum-mechanical proposal for dispersion cancellation [1], a pair of energy-time entangled photons is generated in the source using parametric down-conversion, in which individual photons in a CW [10] pump laser are split into pairs of photons, conserving energy in the process. Photon pairs generated in this way are emitted at very nearly the same time and each photon has a relatively large bandwidth. Nevertheless, the sum of their frequencies is equal to that of the pump laser in the down-conversion source and the two frequencies are anti-correlated. The two photons then propagate through two dispersive media that are separated by a large distance. It was shown in Ref. [1] that the dispersion in one medium can be cancelled out by the dispersion in the other medium if the dispersion coefficients are equal in magnitude but have the opposite sign, and the two photons will still be detected at the same time even though they have both passed through a dispersive medium. The lack of a classical analog for nonlocal dispersion cancellation has already been discussed in Ref. [3].

The difference between the classical model of Ref. [6] and quantum mechanics can be understood by noting that the frequency entanglement of the two beams in the quantum-mechanical case gives an intensity correlation proportional to

$$\langle \hat{I}_1'(t)\hat{I}_2'(t')\rangle = \eta \langle \hat{E}_1^{(-)}(t)\hat{E}_2^{(-)}(t')\hat{E}_1^{(+)}(t)\hat{E}_2^{(+)}(t')\rangle =$$
$$c_f \left| \sum_n c_n \left( e^{-i\omega_0 t} e^{-i\Delta\omega_n t} e^{i\beta_1 \Delta\omega_n^2 L} \right)\left( e^{-i\omega_0 t'} e^{i\Delta\omega_n t'} e^{i\beta_2 \Delta\omega_n^2 L} \right) \right|^2 \quad (20)$$
$$= c_f \left| \sum_n c_n e^{-i\omega_0 (t+t')} e^{i(\beta_1+\beta_2)\Delta\omega_n^2 L} \right|^2$$

where $\eta$ and $c_f$ are constants of no interest here. (More details of this derivation can be found in Ref. [1].)

It can be seen that there is only a single sum over one index $n$ in the quantum-mechanical case, so that the dispersive effects coherently cancel out of the quantum state if $\beta_1 = -\beta_2$. In contrast, there is a double sum in the corresponding classical expression of Eq. (10), in which case the dispersive effects do not cancel out of the fields themselves. As a result, the quantum state is exactly the same after the dispersion as it was before, which is not the case for the classical fields. The single sum in Eq. (20) corresponds to an entangled state, by definition, whereas the classical state of Eq. (10) corresponds to a product state, even if it were to be treated quantum-mechanically. It can be seen from Eq. (10) that the phases of the two classical beams are not really anti-correlated in a product state of that kind, even though Eqs. (1) and (2) suggest that their statistical properties are anti-correlated.

Simply subtracting or multiplying the results of measurements made on two classical signals at distant locations and averaging the results does not make a phenomenon "nonlocal" or demonstrate any physical cancellation, regardless of the correlations that may exist between the two measurements. Nonlocality refers to the fact that, in the quantum-mechanical case, the coherent cancellation of the dispersion coefficients is only possible because the photons are entangled and their properties cannot be described locally. It is well known that classical electromagnetism is a strictly local theory.

Shapiro has argued that the state of the classical fields is the same after they have propagated through the dispersive media [6]. Although their statistical properties are the same, both fields and intensities have changed, as can be seen from Eq. (13). In classical electromagnetism, the field or intensity of a beam can be measured before it enters a dispersive medium without significantly perturbing the system, and that value can be compared with the field leaving the dispersive medium. Those values will not be the same, and Shapiro's claim that the state of the fields has not been changed is incorrect.

The dispersion is clearly not being physically cancelled out in the classical model, but one might ask whether or not it is really cancelled out in the quantum-mechanical case or if the photons can be interpreted as undergoing identical dispersion as well. From a quantum-mechanical perspective, the dispersion coefficients literally cancel out of Eq. (20) and the state of both photons is unchanged by the dispersive media. The only reasonable classical explanation would be that short pulses are being emitted at random but correlated times, and that the widths of these pulses are somehow not affected due to the combined or cooperative effects of the dispersive media.

The situation here is different from the local form of dispersion cancellation suggested by Steinberg et al. [11,12], in which two entangled photons are recombined to interfere on a single beam splitter. The local nature of this process allows two classical fields to interfere and a physical cancellation of dispersion can occur classically, as has been shown using sum-frequency generation [13]. The nonlocal nature of the dispersion cancellation predicted in Ref. [1], in which the two photons never return to the same location, prevents any true cancellation of dispersion from occurring in a classical model [3].

## V. SUMMARY AND CONCLUSIONS

Shapiro has considered two "phase-sensitive" classical beams of light whose phases and frequencies are anti-correlated, and he has shown that the cross-correlation of their intensities (the usual peak in the Hanbury-Brown and Twiss effect) is unaffected by two dispersive media with equal and opposite dispersion coefficients [6]. This does not correspond to any physical cancellation of the dispersion in the two beams, and it was shown here that both beams undergo identical dispersion in such a way that statistical properties, such as the difference between the two intensities, are unaltered. This situation is completely different from the nonlocal cancellation of dispersion for entangled pairs of photons, where the dispersion itself is eliminated as a result of nonlocal interference effects.

The origin of these features in the correlation functions was shown to be totally different in the classical and quantum-mechanical cases, contrary to Shapiro's suggestion that "their physical origins are identical and essentially classical". The classical derivation is based on Eq. (10) while the quantum-mechanical derivation is based on Eq. (20). These two equations are totally different in form and the origin of the effects is obviously different from a mathematical point of view. Physically, the two effects are fundamentally different in that Eq. (20) corresponds to a single sum over the modes of the field which allows a coherent cancellation of the dispersion coefficients and the total elimination of the effects of dispersion. In contrast, the classical Eq. (10) contains a double sum over the modes of the field at the two locations, which corresponds to a product state that results in local dispersion in both beams. The same dispersion occurs in both beams so that the cross-correlation of their intensities is unaffected without any physical cancellation of the dispersion.

More importantly, the essential feature of nonlocal dispersion cancellation is the fact *all* of the photon pairs remain coincident after they have traversed two distant dispersive media [1]. Shapiro's model does not reflect this aspect of nonlocal dispersion cancellation. Although there are some detection events that remain coincident even in the presence of the dispersive media, there are many more detection events that occur at uncorrelated times, as illustrated in Fig. 2. In a classical model, the only way to produce detection events that are all coincident, as they are in the quantum-mechanical case, would be to emit very short pulses at correlated times. But the width of the classical pulses would be greatly increased due to dispersion in the two media, regardless of the sign of the dispersion coefficients, and such a classical model cannot reproduce or explain the fundamental feature of nonlocal dispersion cancellation that makes it nonclassical. As a result, it seems apparent that there really is no classical analog for nonlocal dispersion cancellation [3].

## ACKNOWLEDGEMENTS

I would like to thank J.H Shapiro for providing an advance copy of his paper and for interesting discussions. This work was supported in part by the National Science Foundation (NSF) under grant 0652560.